\documentclass{ethpaper}
\usepackage{graphicx}

\begin{document}
\begin{titlepage}
\ethnote{}
\title{Studies of the effect of charged hadrons on lead tungstate crystals}
\begin{Authlist}
Francesca Nessi-Tedaldi

\Instfoot{eth}{Swiss Federal Institute of Technology (ETH), CH-8093 Z\"urich, Switzerland}
\end{Authlist}
\maketitle

\begin{abstract}
  Scintillating crystals are used for calorimetry in several
  high-energy physics experiments.  For some of them, performance has
  to be ensured in difficult operating conditions, like a high
  radiation environment, very large particle fluxes and high collision
  rates.  Results are presented here from a thorough series of
  measurements concerning mainly the effect of charged hadrons on lead
  tungstate. It is also shown how these results can be used to predict
  the effect on crystals due to a given flux of particles.
\end{abstract}

\vspace{7cm}
\conference{\em Submitted to Proceedings Calor 2008 - XIII International Conference on Calorimetry in High Energy Physics\\ Pavia (Italy) 26-30 May 2008\\ to be published in Journal of Physics: Conference Series}

\end{titlepage}

\section{Introduction}

The effect of large hadron fluxes on $\mathrm{PbWO}_4$ crystals has
become an important subject of investigation with the construction of
the Large Hadron Collider (LHC) at CERN and of experiments using such
crystals. The question had to be studied, whether large hadron
fluences cause a specific, possibly cumulative damage, and if so, what
its quantitative importance is and whether it only affects light
transmission or also the scintillation properties of lead tungstate.
It should be noticed that all damage observed in earlier tests
\cite{r-BAT1,r-AUF} could always be ascribed to the ionising dose
associated with the hadron flux, apart from some indication of a
hadron-specific damage in BGO, which can be extracted \cite{r-FN2}
from existing data, as discussed later herein. However, none of the
previous tests on lead tungstate had been extended to the full
integrated fluences expected at the LHC.  Therefore, a thorough series
of irradiation tests on lead tungstate crystals
\cite{r-LTNIM,r-LYNIM,r-PIONNIM} has been performed over several
years, with exposures reaching up to the highest integrated fluence
expected at the LHC after 10 years of running.  In this presentation,
an overview is given on the main results obtained through these tests,
and the understanding reached is summarised.

\section{Proton and \mbox{\boldmath$\gamma$} irradiation studies}
In a first study to explore the dependence on the integrated fluence
of a possible hadron-specific damage \cite{r-LTNIM}, eight CMS
\cite{r-TDR} production crystals of consistent quality \cite{r-TDR}
were irradiated in a 20 to 24 GeV/c proton flux of either
$\phi_p=10^{12}\;\mathrm{p\; cm}^{-2}\mathrm{h}^{-1}$ or
$\phi_p=10^{13}\;\mathrm{p\; cm}^{-2}\mathrm{h}^{-1}$ at the CERN PS
accelerator \cite{r-IR1}. The maximum fluence reached was $\Phi_p =
5.4 \times 10^{13}\;\mathrm{p\; cm}^{-2}$. The crystals were nearly
parallelepipedic, with dimensions $2.4\times 2.4\times 23\;
\mathrm{cm}^3$.  To disentangle the contribution to damage from the
associated ionising dose, complementary $^{60}\mathrm{Co}\;
\gamma$-irradiations were performed at a dose rate of 1 kGy/h on
further seven crystals, since a flux $\phi_p = 10^{12}\;\mathrm{p\;
  cm}^{-2}\mathrm{h}^{-1}$ has an associated ionising dose rate in
$\mathrm{PbWO}_4$ of 1 kGy/h. The maximum dose reached, of $50.3$ kGy,
is just $\sim 50\%$ below the one reached in proton irradiations.
\begin{figure}[b]
\begin{minipage}{20pc}
\includegraphics[width=20pc]{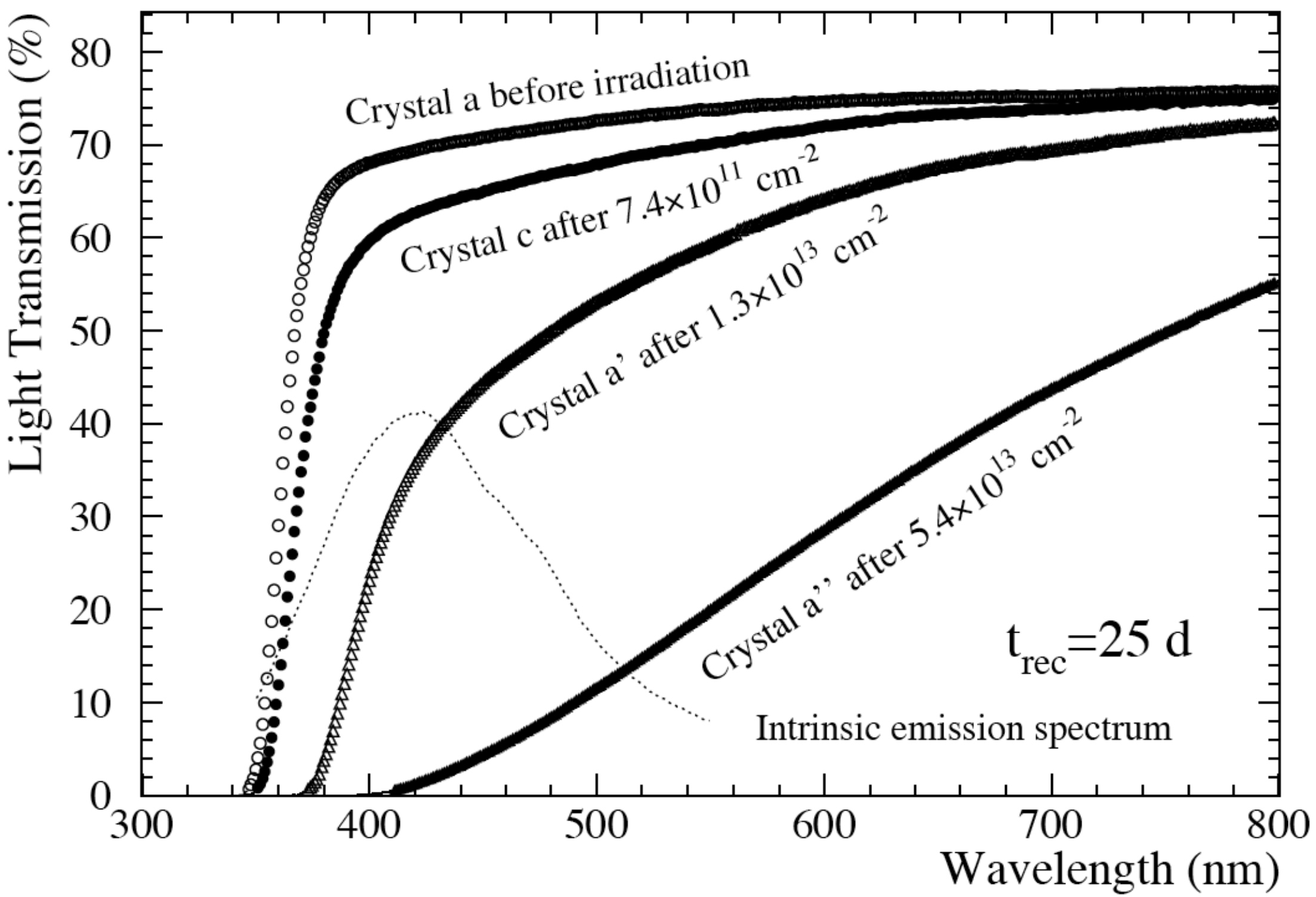}
\caption{\label{f-LTp}Light transmission curves for crystals with various
degrees of proton-induced radiation damage \cite{r-LTNIM}. The emission
spectrum units \cite{r-ZHUIEEE} are arbitrary.}
\end{minipage}\hspace{2pc}%
\begin{minipage}{15pc}
\includegraphics[width=15pc]{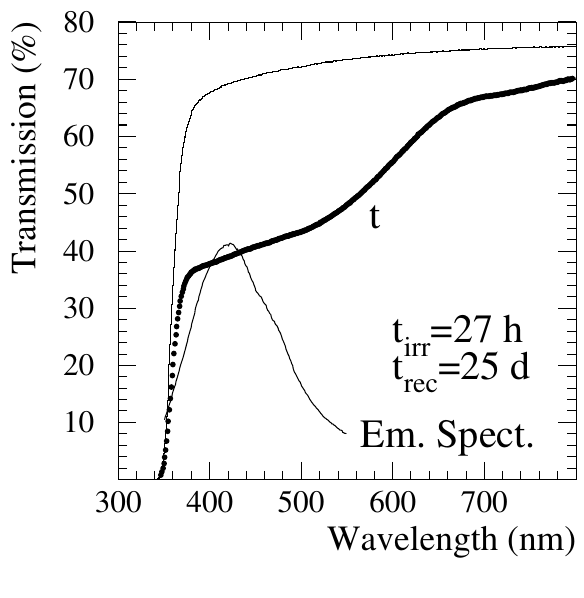}
\caption{\label{f-LTg}Transmission curves for a $\gamma$-irradiated
crystal \cite{r-LTNIM} prior to (thin)
 and after (thick line) irradiation. Emission spectrum as in
figure \ref{f-LTp}.}
\end{minipage} 
\end{figure}
The measurements of transmission damage revealed \cite{r-LTNIM} that
proton irradiation decreases the light transmission for all
wavelengths and moves the ultra-violet transmission band-edge to
longer wavelengths, as can be seen in figure \ref{f-LTp}.  In
$\gamma$-irradiated crystals, the transmission band-edge does not
shift at all, even after the highest integrated dose: one only
observes the well-known absorption band \cite{r-ZHU1} around 420 nm
that can be appreciated in figure \ref{f-LTg}. These results
demonstrate the qualitatively different nature of proton-induced and
$\gamma$-induced damage.
\begin{figure}[t]
 \begin{center}
\includegraphics[width=38pc]{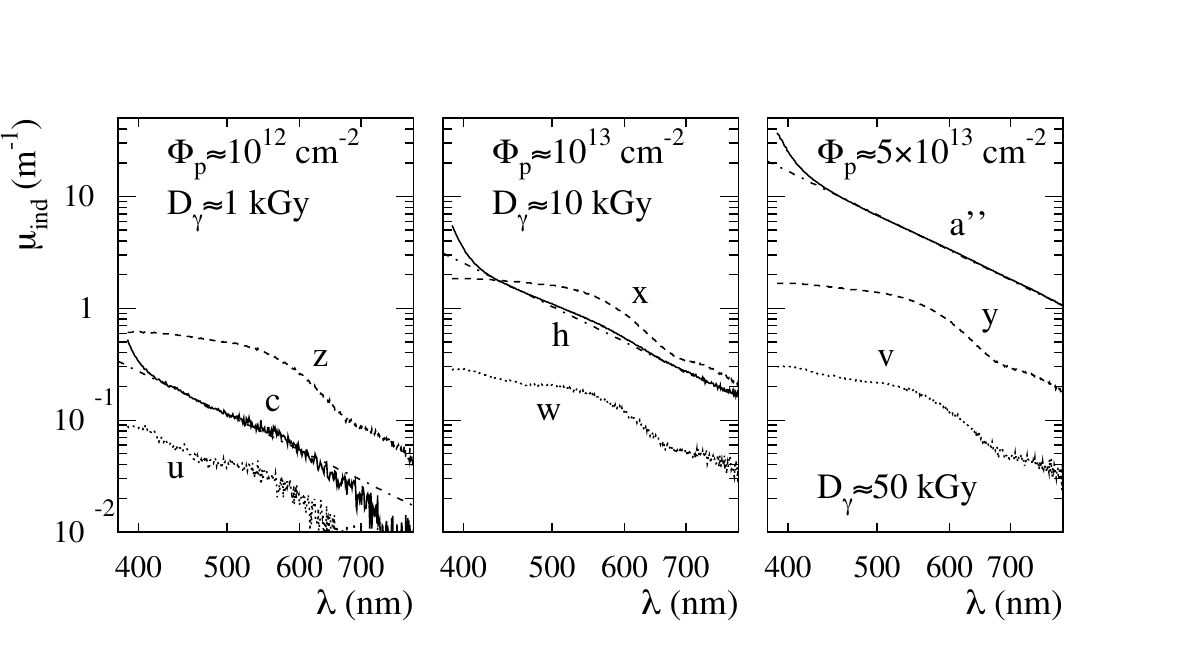}
\caption{\label{f-RAYL}Plot of $\mu_{IND}(\lambda)$ versus $\lambda$ for proton-irradiated crystals ($a''$, $c$, $h$) and for
$\gamma$-irradiated ones ($u$, $v$, $w$, $x$, $y$, $z$) \cite{r-LTNIM}. The dot-dashed line shows $\lambda^{-4}$, fitted to the proton-damage data.}
\end{center}
\end{figure}
\begin{figure}[b]
\begin{minipage}{18pc}
\begin{center}
\includegraphics[width=18pc]{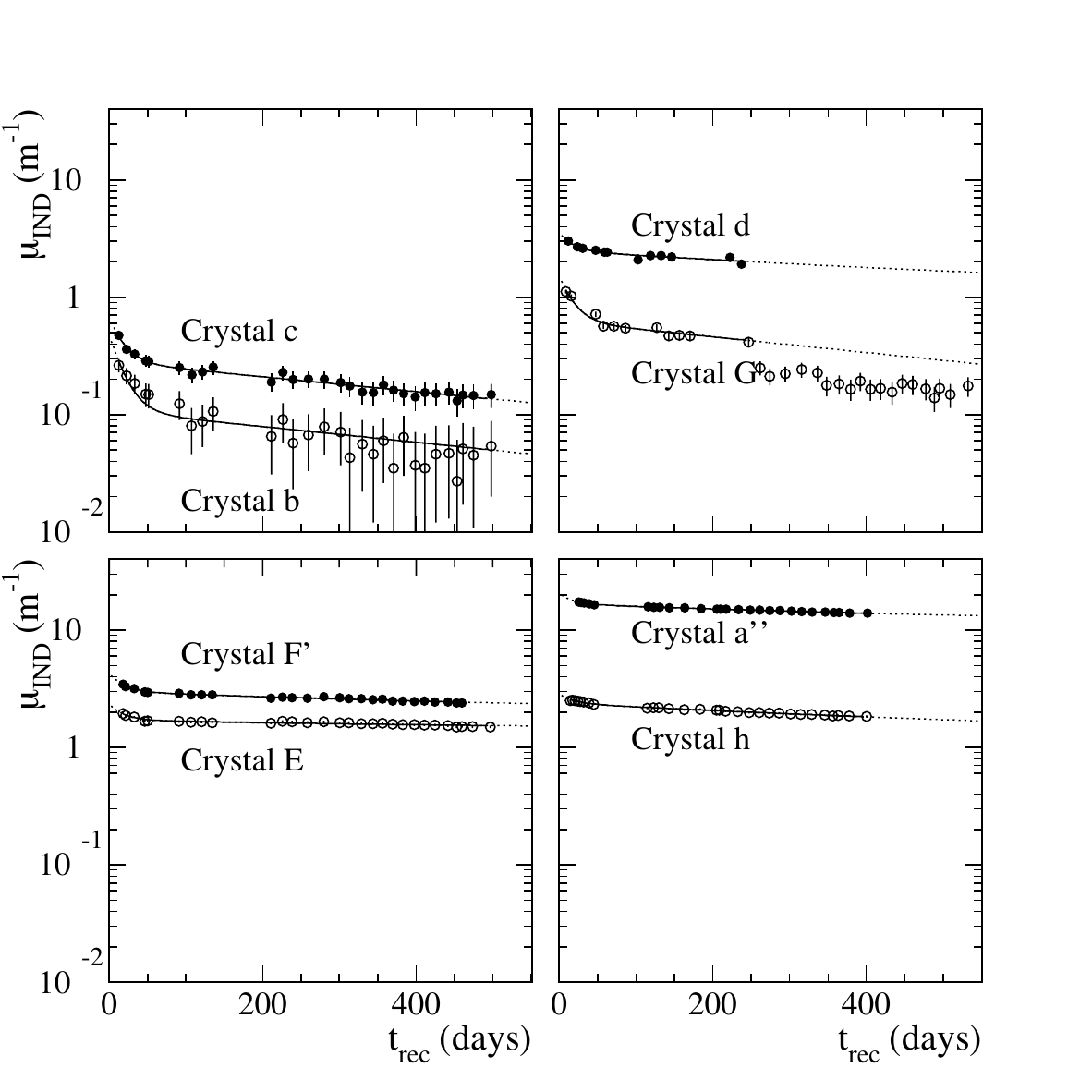}
\end{center}
\caption{\label{f-pREC}Evolution of longitudinal induced absorption
$\mu^{LT}_{IND}(420 \; \mathrm{nm})$ over time for proton irradiated
crystals \cite{r-LTNIM}.}
\end{minipage}\hspace{2pc}%
\begin{minipage}{18pc}
\begin{center}
\includegraphics[width=18pc]{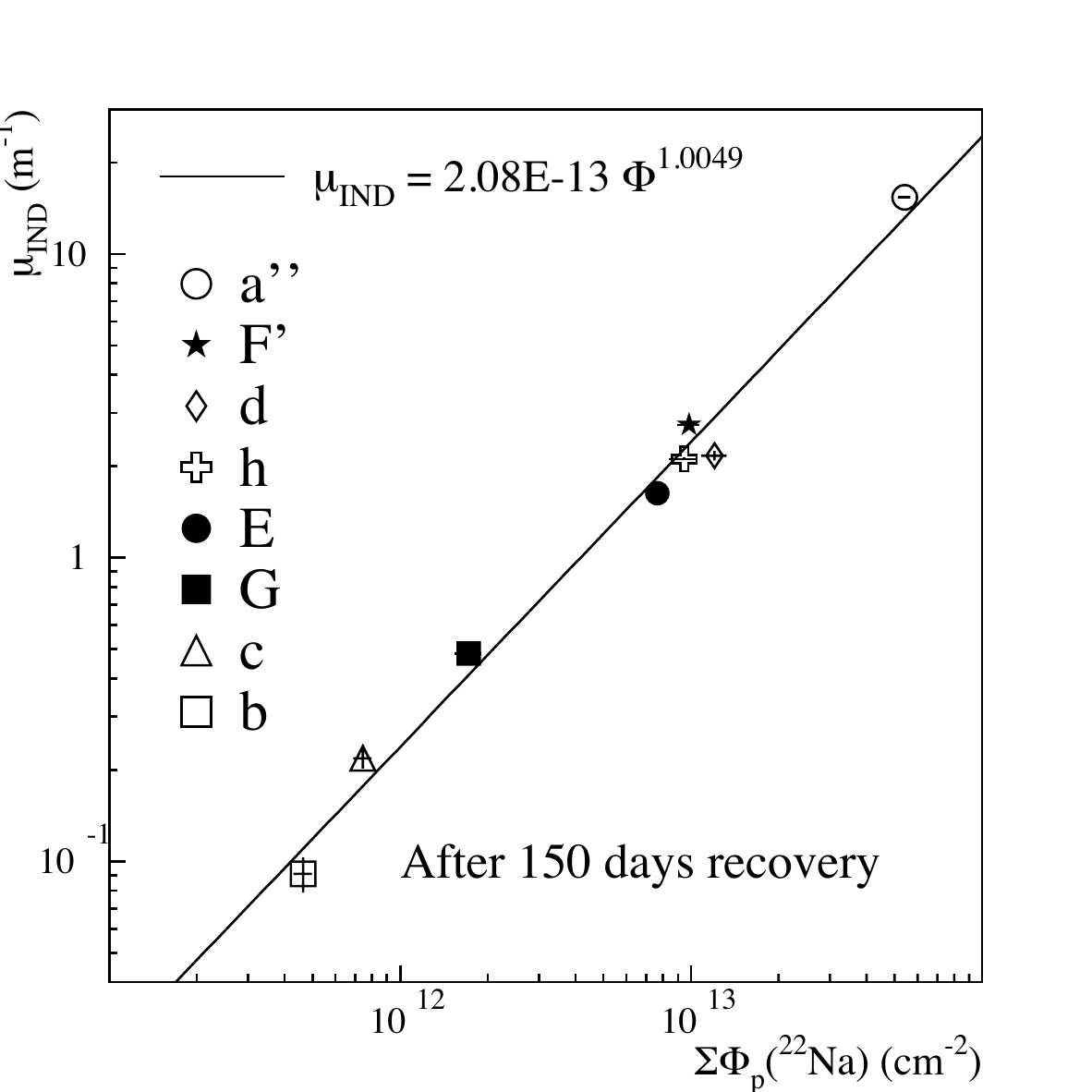}
\end{center}
\caption{\label{f-LIN}Induced absorption $\mu_{IND}(420 \; \mathrm{nm})$ as
a function of cumulative proton
fluence \cite{r-LTNIM}.}
\end{minipage} 
\end{figure}
\begin{figure}[h]
\begin{minipage}{18pc}
\begin{center}
\includegraphics[width=18pc]{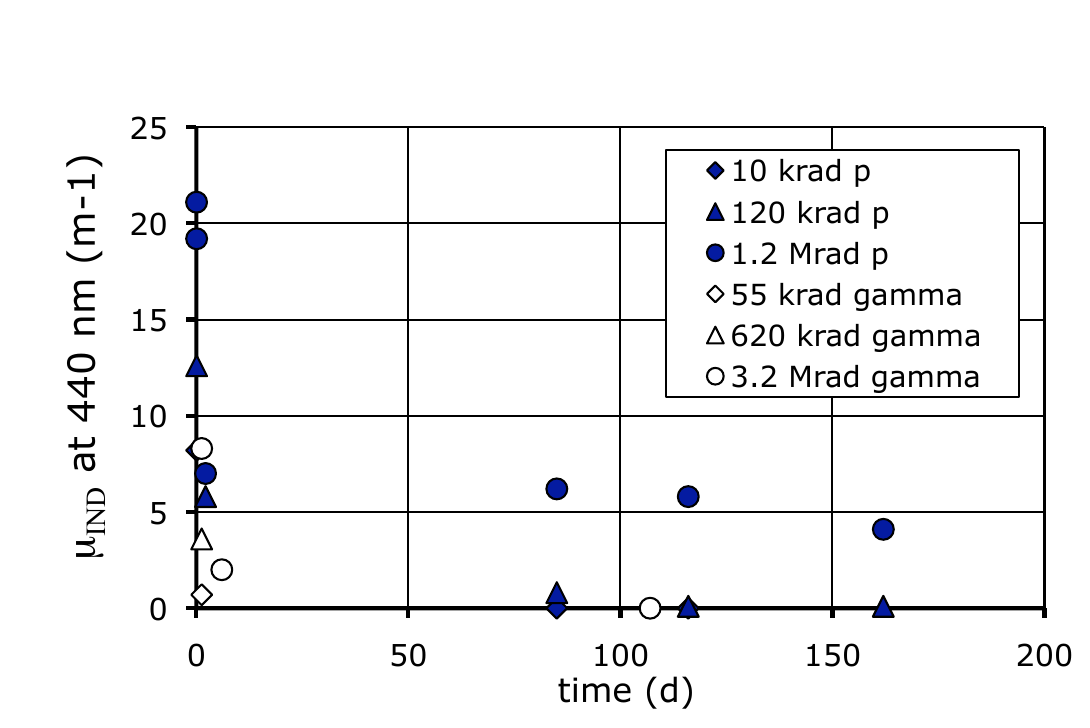}
\end{center}
\caption{\label{f-RECBGO} Evolution over time of the Longitudinal induced
absorption, $\mu_{IND}(440\;\mathrm{nm})$,  for proton- (full symbols) and
for $\gamma$-irradiated (white symbols) BGO crystals, extracted from data
in \cite{r-KobaBGO}.}
\end{minipage}\hspace{2pc}%
\begin{minipage}{18pc}
\begin{center}
\includegraphics[width=17pc]{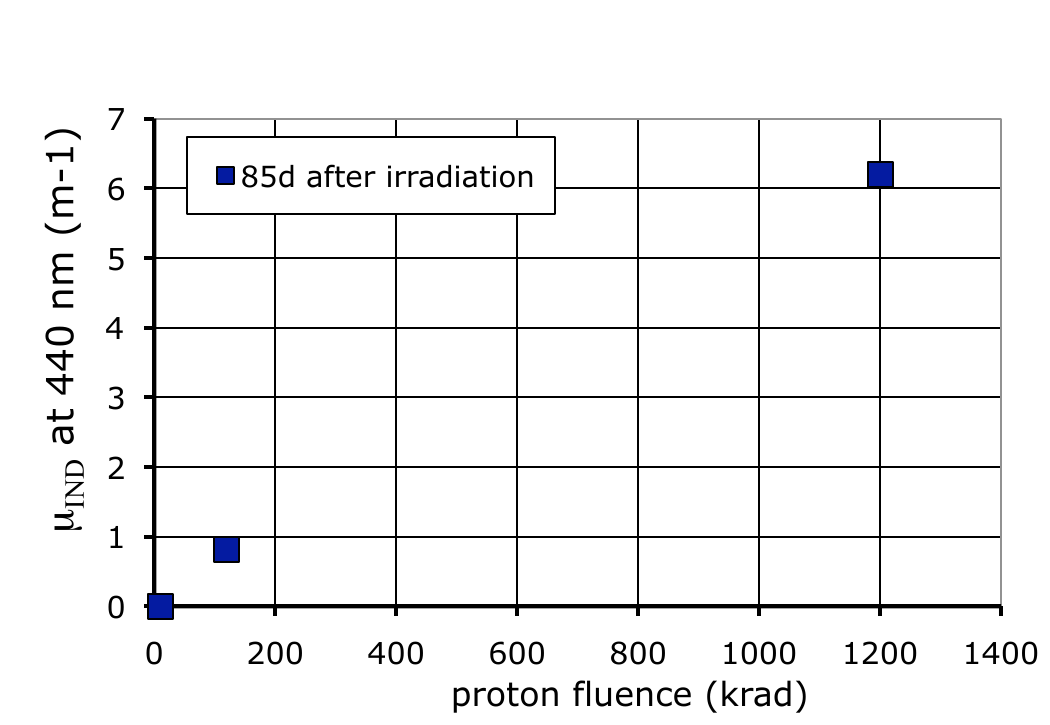}
\end{center}
\caption{\label{f-LINBGO}Correlation between the absorption
$\mu_{IND}(440\;\mathrm{nm})$ induced by proton irradiation and proton
fluence in BGO, extracted from data in \cite{r-KobaBGO}.}
\end{minipage} 
\end{figure}
\begin{figure}[b]
\begin{minipage}{18pc}
\begin{center}
\includegraphics[width=18pc]{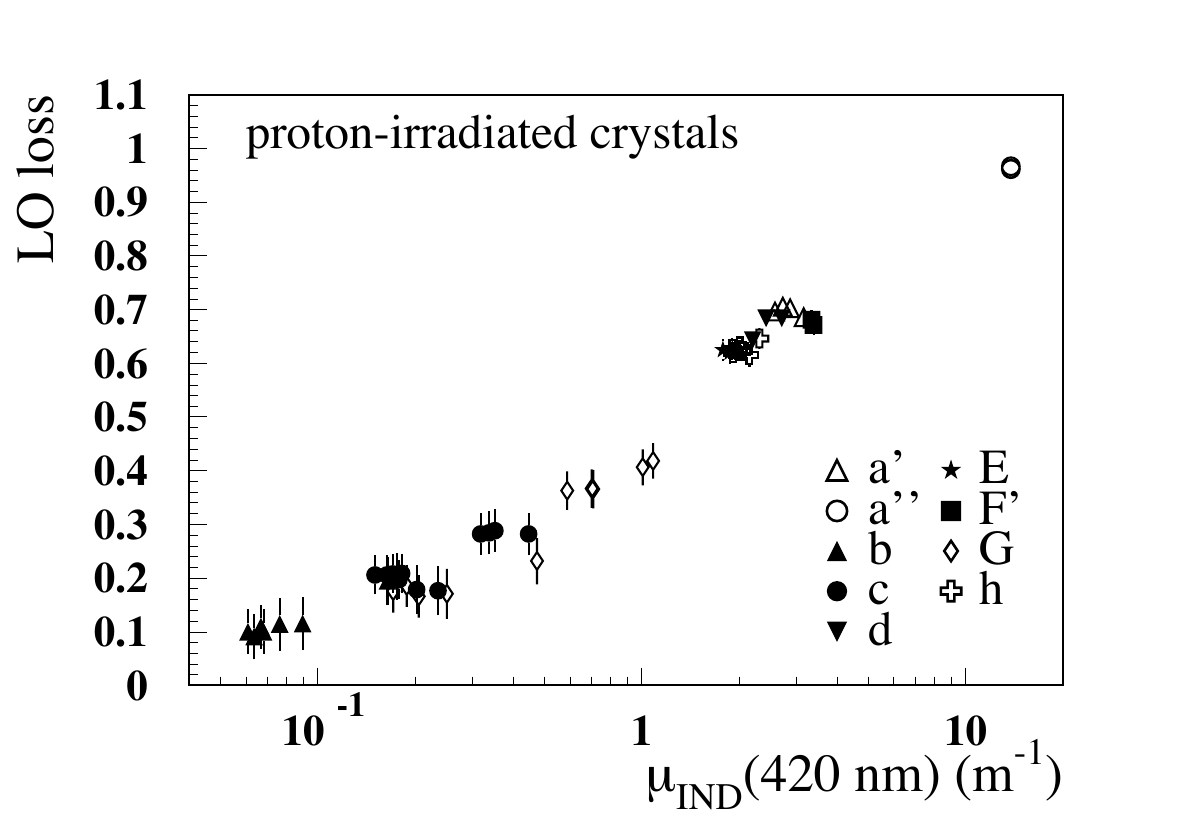}
\end{center}
\caption{\label{f-CORRELp}Correlation between induced absorption,
$\mu_{IND}(420 \; \mathrm{nm})$, and
Light Output loss for proton induced damage in $\mathrm{PbWO}_4$
\cite{r-LYNIM}.}
\end{minipage}\hspace{2pc}%
\begin{minipage}{18pc}
\begin{center}
\includegraphics[width=18pc]{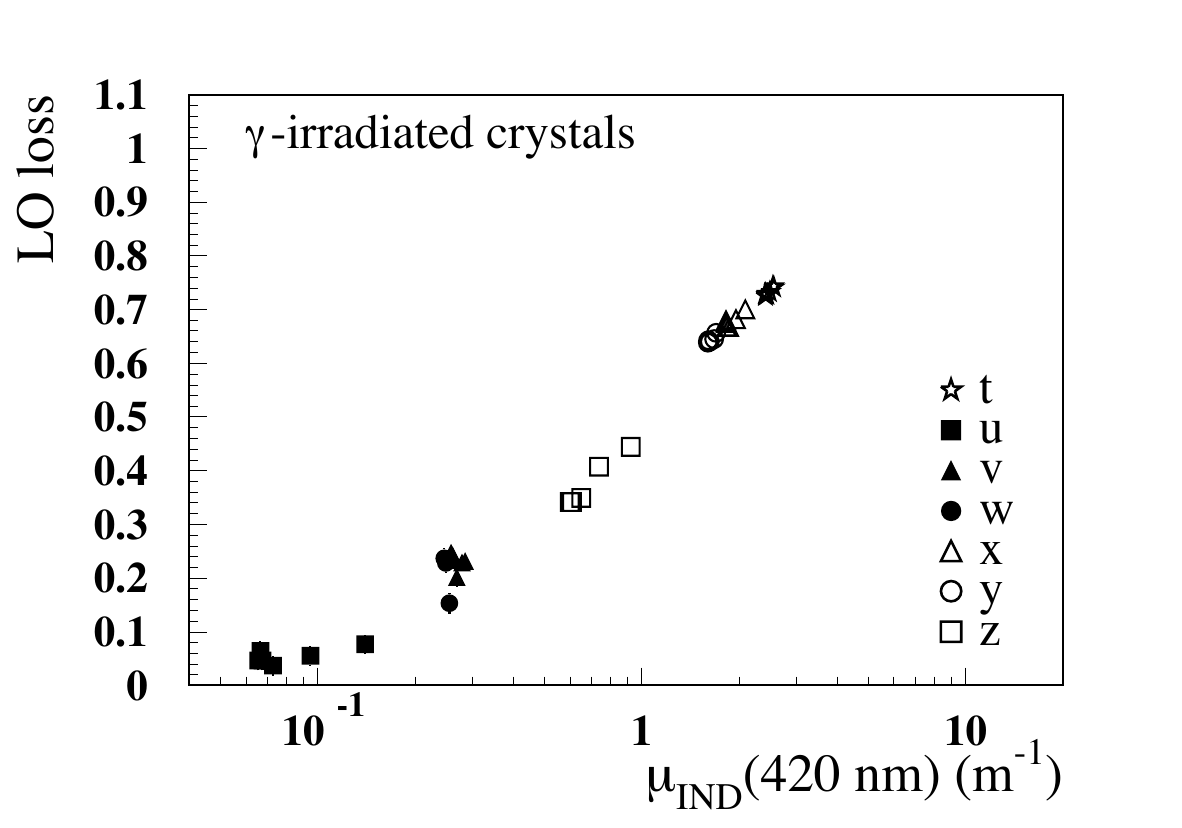}
\end{center}
\caption{\label{f-CORRELg}Correlation between induced absorption,
$\mu_{IND}(420 \; \mathrm{nm})$, and
Light Output loss for $\gamma$-induced damage in $\mathrm{PbWO}_4$
\cite{r-LYNIM}.}
\end{minipage} 
\end{figure}

The damage is usually quantified through the induced absorption
coefficient for a given wavelength $\lambda$, which for the
Longitudinal Transmission, measured through the length $\ell$ of a
crystal, is defined as
\begin{equation}
\mu_{IND}^{LT}(\lambda) = \frac{1}{\ell}\times \ln \frac{LT_0}{LT}
\label{muDEF}
\end{equation}
where $LT_0\; (LT)$ is the Longitudinal Transmission value measured
before (after) irradiation through the length of the crystal, and
analogously for Transverse Transmission (TT).  The data for
$\mu^{LT}_{IND}$ versus light wavelength for proton-irradiated
crystals in figure \ref{f-RAYL} show a $\lambda^{-4}$ dependence,
which is absent in $\gamma$ - irradiated crystals.  This is an
indication of Rayleigh scattering from small centres of severe damage,
as they might be caused by the high energy deposition of heavily
ionising fragments along their path, locally changing the crystal
structure.  Since the crystals contain heavy elements, Pb and W, it
was argued in \cite{r-LTNIM} that a hadron-specific damage can in fact
arise from the production, above a $\sim$20 MeV threshold, of heavy
fragments with up to 10 $\mu$m range and energies up to $\sim$100 MeV,
causing a displacement of lattice atoms and energy losses, along their
path, up to 50000 times the one of minimum-ionising particles. The
observed behaviour confirms this qualitative understanding. A further
evidence evidence of Rayleigh scattering is the complete polarisation
of scattered light, as observed using Polaroid films.

As it can be seen in figure \ref{f-pREC}, where the evolution over
time of damage values measured at the peak of scintillation emission
wavelength, $\mu^{LT}_{IND}\mathrm{(420\; nm)}$, is plotted, only a
small fraction of the hadron damage recovers on a time scale
comparable to the expected duration of LHC running.

The correlation between the induced absorption coefficient at
420 nm and fluence \cite{r-LTNIM} shown in figure \ref{f-LIN}, is
consistent with a linear behaviour over two orders of magnitude,
showing that proton-induced damage in $\mathrm{PbWO}_4$ is
predominantly cumulative, unlike $\gamma$-induced damage, which
reaches equilibrium \cite{r-LTNIM,r-ZHU1}. No flux dependence was
observed.

Damage values from protons and $\gamma$'s were also compared in a
study performed on BGO\cite{r-KobaBGO}. The changes in band-edge are
similar to what is seen in $\mathrm{PbWO}_4$, and long enough after
irradiation, when the ionising-radiation damage contribution has
recovered (figure \ref{f-RECBGO}), a remaining proton-induced damage
can be extracted, that behaves linearly with fluence, as visible in
figure \ref{f-LINBGO}.
 
The relevant quantity for detector operation being the scintillation
light output, any effect of hadrons on this quantity is of great
importance. Detector calibration through a light injection system, as
foreseen e.g. by CMS \cite{r-TDR}, is based on the assumption that
changes in light output are all due to changes in light transmission.
The study published in \cite{r-LYNIM} compares the correlations
between Light Output loss and induced absorption in proton- and
$\gamma$-irradiations for the same set of crystals above. Those
correlations (figures \ref{f-CORRELp} and \ref{f-CORRELg}) show that,
in the explored range of proton fluences, no additional,
hadron-specific damage to the scintillation mechanism could be
observed in lead tungstate within the measurement's precision.

\section{Comparative proton and positive §pion irradiation studies}
Crystals used in high-energy physics detectors will typically be
exposed to hadrons -- mostly charged pions -- with different
energies. In the CMS experiment at the LHC for example \cite{r-TDR},
the large hadron fluxes are due to particles - mostly pions - whose
energies rarely exceed 1 GeV.  To understand how to scale
proton-damage values to the ones caused by such pion fluences, a
previously $\gamma$ irradiated crystal ($w$ in \cite{r-LTNIM} and in
figure \ref{f-RAYL}) was cut into three, $7.5$ cm long, sections
($w1$, $w2$ and $w3$) after thermal annealing of its $\gamma$-induced
damage.

The middle section, $w2$, was irradiated with 290 MeV/c positive pions
at the Paul Scherrer Institute in Villigen, Switzerland.  Details
about the irradiation are found in \cite{r-PIONNIM}.  The crystal was
irradiated up to a total fluence $\Phi_{\pi}=(5.67 \pm 0.46)\times
10^{13}\; \mathrm{cm^{-2}}$, while the average flux on the crystal was
$\phi_{\pi}=4.13\times 10^{11}\; \mathrm{cm^{-2}\; h^{-1}}$.

\begin{figure}[h]
\includegraphics[width=19pc]{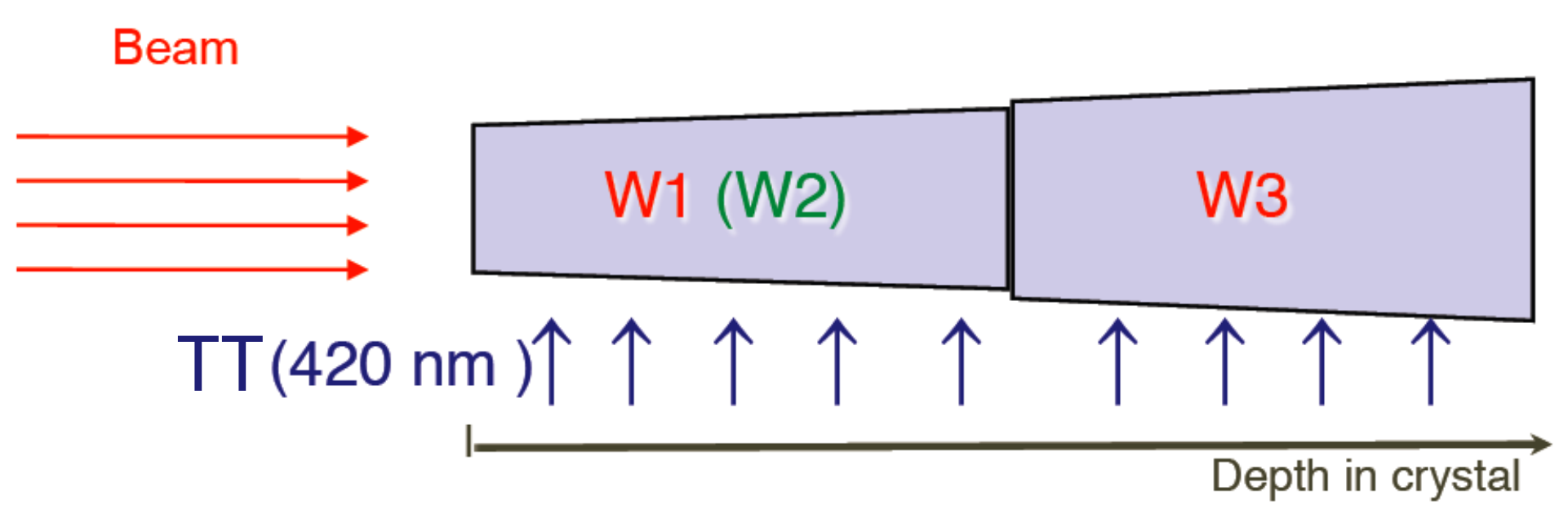}\hspace{2pc}
\begin{minipage}[b]{16pc}\caption{\label{f-SKETCH} Illustration of $w1$
and $w3$ ($w2$) crystal placement during proton (pion) irradiation, and
of the TT measurement direction after
irradiation.}
\end{minipage}%
\end{figure}

The first, {\em w1}, and last, {\em w3}, sections were irradiated with
24\,GeV/c protons. They were placed at the same time, $w1$ in front of
$w3$, into the proton-irradiation facility \cite{r-IR1} of the CERN PS
accelerator, with the beam entering through the small {\em w1} face,
so that the hadronic cascade could develop through both crystals, as
illustrated in figure \ref{f-SKETCH}.  The procedure was totally
analogous to the one described in \cite{r-LTNIM}.  The proton fluence
on the $w1$ crystal front face was $\Phi_p=(1.17 \pm 0.11) \times
10^{13}\;\mathrm{cm}^{-2}$.
\begin{figure}[b]
\includegraphics[width=20pc]{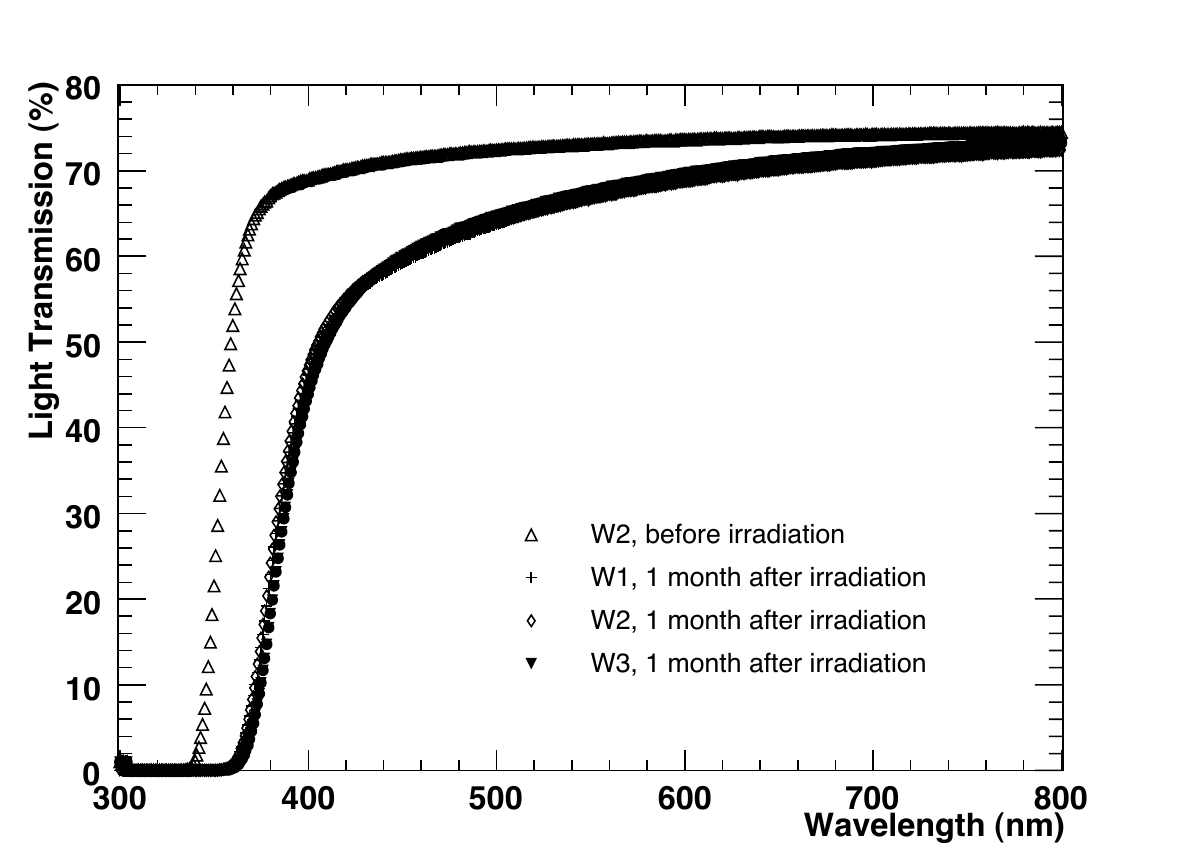}\hspace{2pc}%
\begin{minipage}[b]{15pc}\caption{\label{f-LTpion}Longitudinal light
transmission curves for the three
sections of crystal $w$, one month after hadron irradiation, in
comparison to the values before irradiation \cite{r-PIONNIM}.}
\end{minipage}
\end{figure}
\begin{figure}[t]
\begin{center}
\includegraphics[width=38pc]{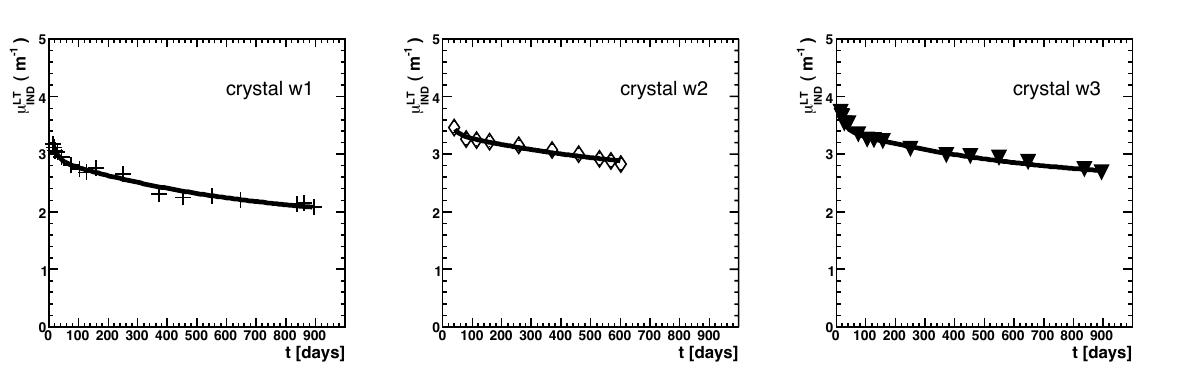}
\end{center}
\caption{\label{f-PIONREC}Evolution of the longitudinal induced
absorption $\mu^{LT}_{IND}(420 \; \mathrm{nm})$ over time, showing
recovery, for the three sections of crystal $w$ \cite{r-PIONNIM}.}
\end{figure}
\begin{figure}[b]
\begin{minipage}[b]{16pc}
\caption{\label{f-TT}Transverse induced absorption, $\mu_{IND}^{TT}$(420 nm)
as a function of position
along the crystal \cite{r-PIONNIM}. Data for $w1$ and $w3$ are shown
according to the crystals'
position during irradiation.}
\end{minipage} 
\includegraphics[width=22pc]{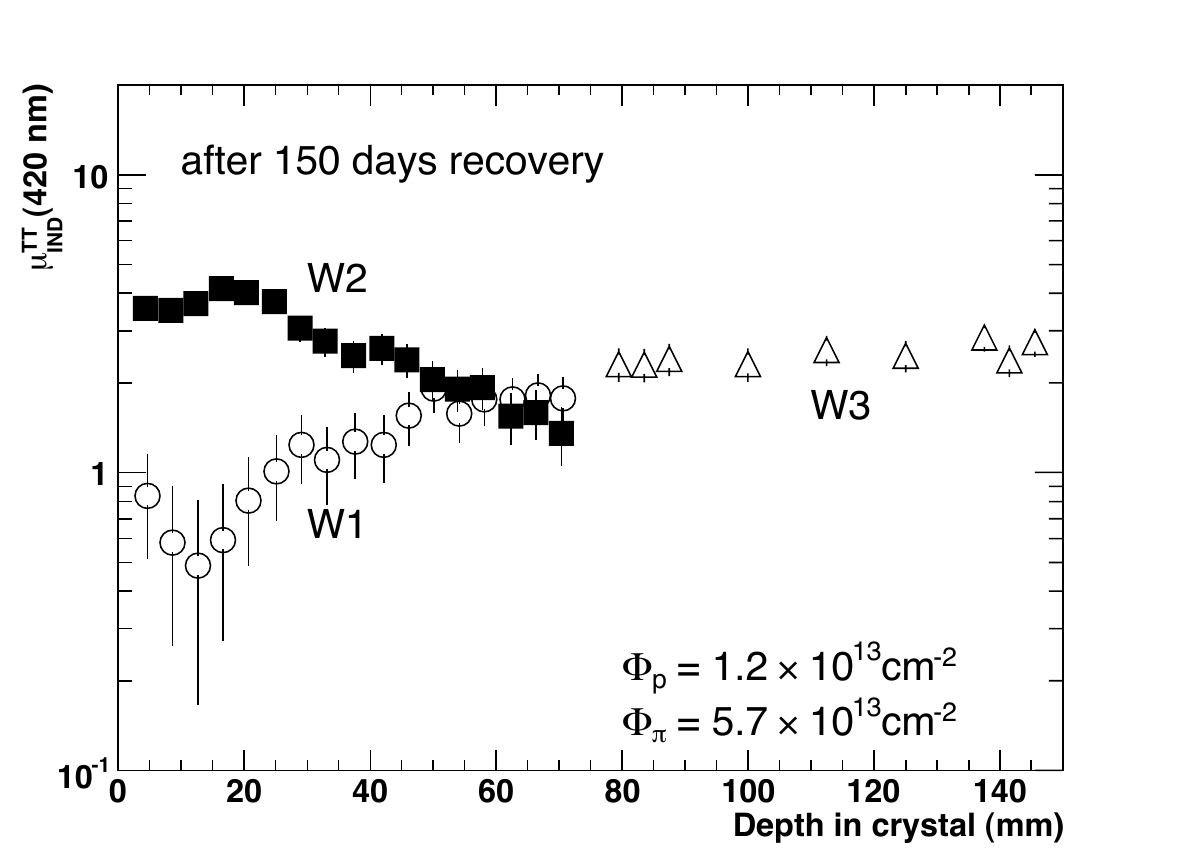}
\end{figure}
\begin{figure}[t]
\begin{minipage}{19pc}
 \includegraphics[width=19pc]{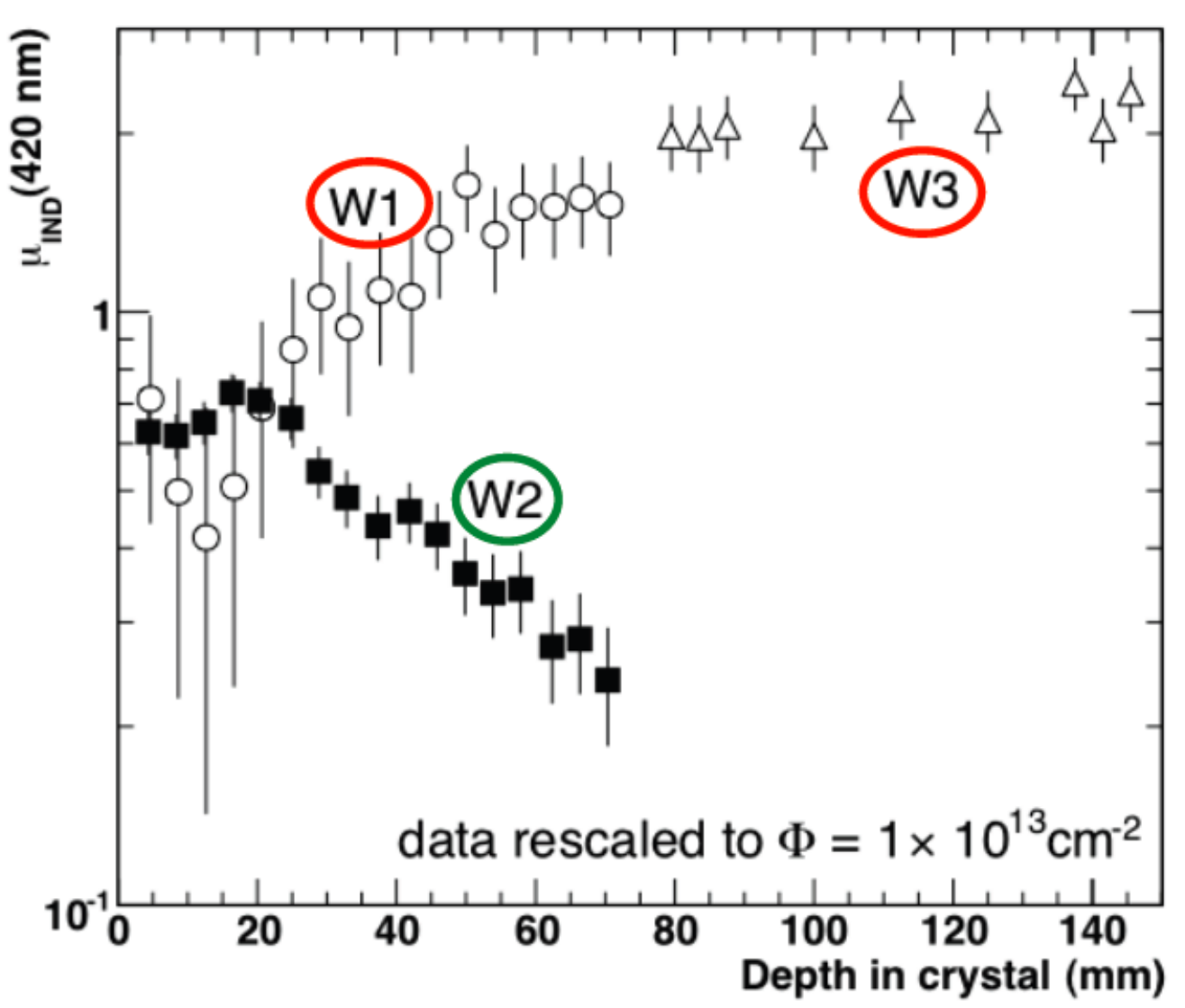}
\caption{\label{f-NORMMU}Same data as in figure \ref{f-TT}, rescaled
to a fluence of  $10^{13}\mathrm{cm}^{-2}$.}
\end{minipage}\hspace{2pc}%
\begin{minipage}{16.5pc}
\includegraphics[width=16.5pc]{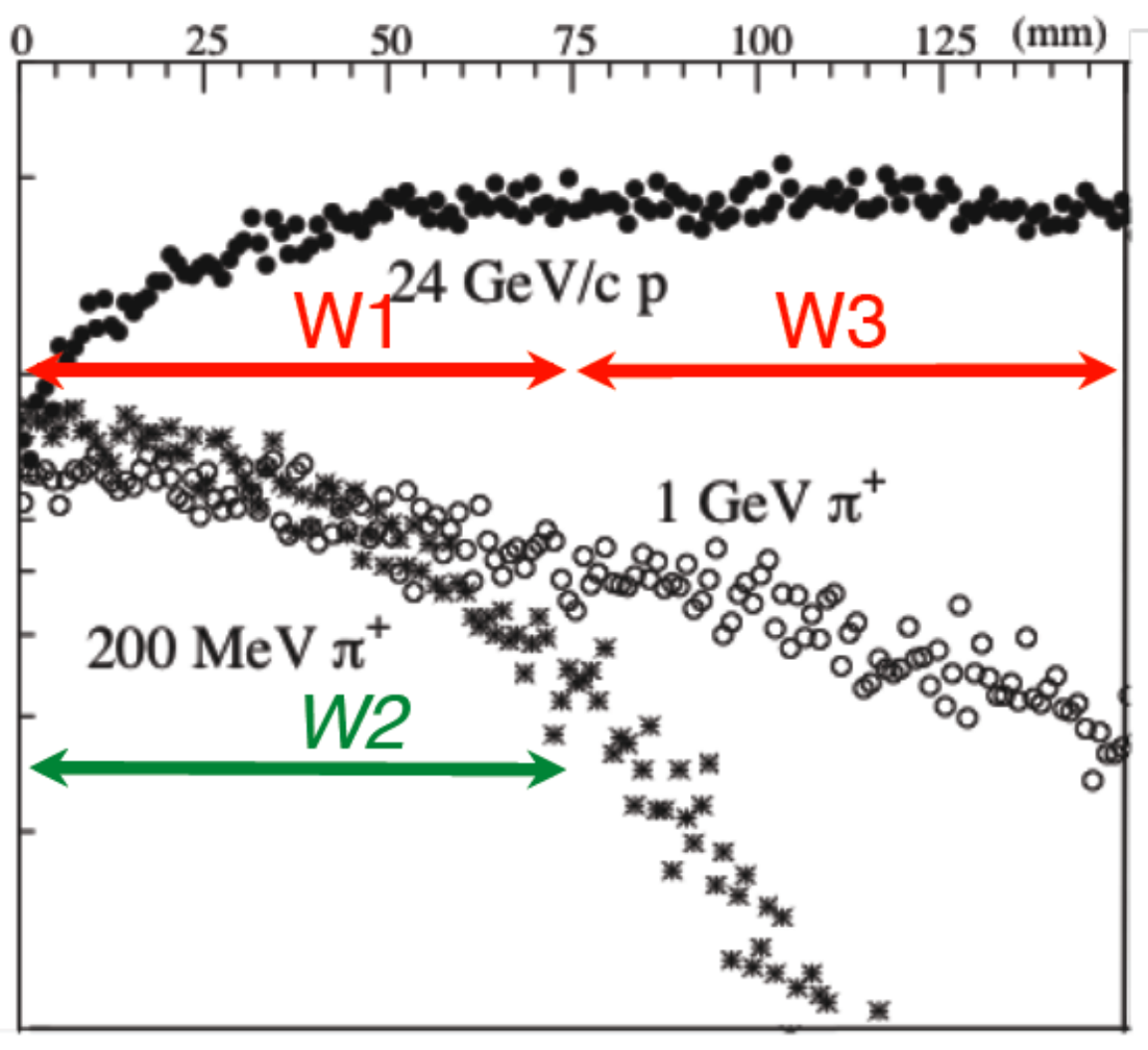}
\caption{\label{f-STARRHO}Star densities from FLUKA Monte Carlo
simulations, as a function of depth in the crystal
\cite{r-LTNIM},\cite{r-HUH}.}
\end{minipage} 
\end{figure}
\begin{figure}[t]
\includegraphics[width=20pc]{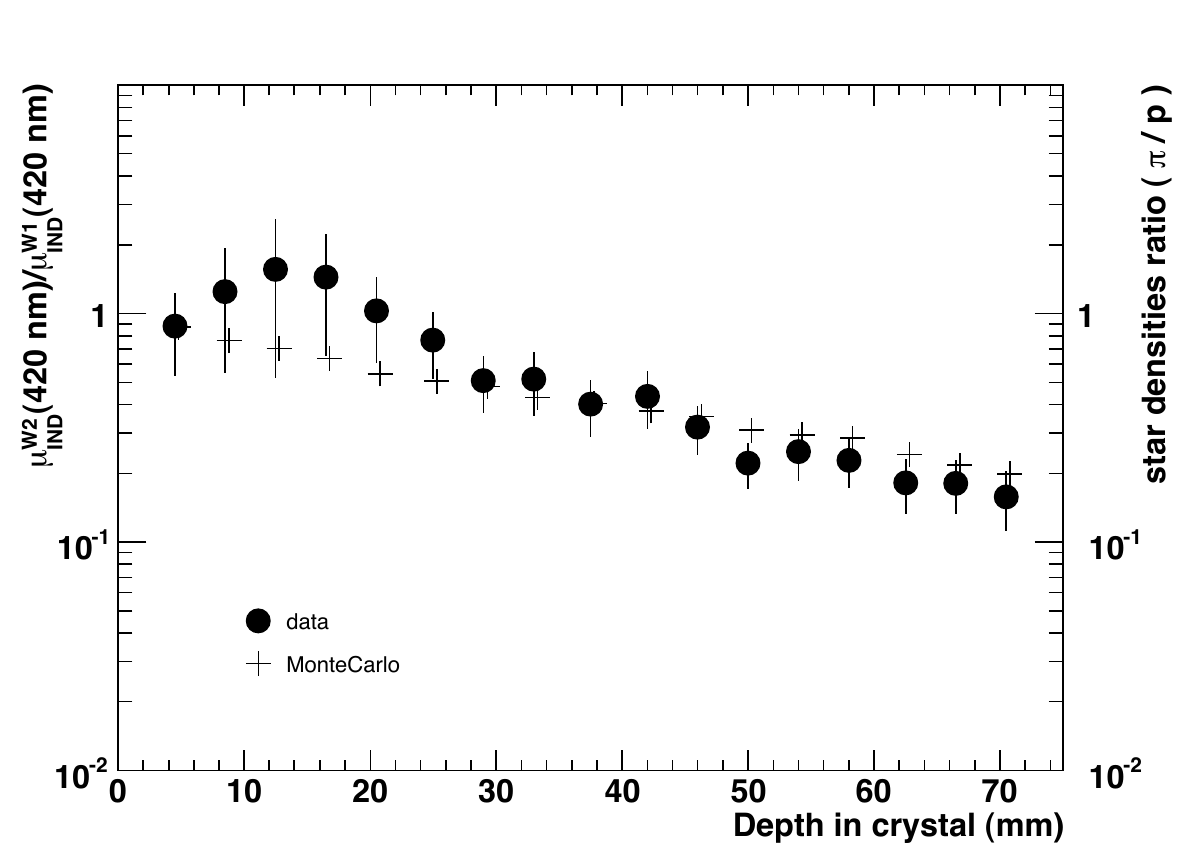}\hspace{2pc}%
\begin{minipage}[b]{15pc}\caption{\label{f-RATIOS}Ratio of
$\mu_{IND}^{TT}$(420 nm) results for $w2$ and $w1$ crystals from
figure \ref{f-NORMMU}, compared to the corresponding simulated star
density ratios from figure \ref{f-STARRHO} \cite{r-PIONNIM}.}
\end{minipage}
\end{figure}
 
After it was demonstrated, in \cite{r-LYNIM}, that hadron damage only
affects light transmission in $\mathrm{PbWO_4}$, the study of pion
damage was focussed on this observable.

At first sight (figure \ref{f-LTpion}) the light transmission measured
through the length of the crystal as a function of wavelength shows,
after irradiation, the same behaviour for positive pions and protons.
The band-edge shift is observed, which is characteristic for
hadron-irradiated lead tungstate and BGO. It should be pointed out
that similar damage amplitudes for the three crystals were obtained on
purpose through a suitable choice of fluences.

The evolution of damage with time for all three crystals is shown in
figure \ref{f-PIONREC}, where it is evident that pion damage follows
the same pattern as observed for proton-irradiated crystals
\cite{r-LTNIM}. The data, as indicated by the lines, are well fitted
by the same function used in \cite{r-LTNIM}, namely a sum of a
constant plus two exponentials with time constants $\tau_1 = 17.2$
days and $\tau_2 = 650$ days. A similarly good fit is obtained through
the form
\begin{equation}
\mu_{IND}^{LT,\; j}(420\; \mathrm{nm},t_{\rm{rec}}) =
                  B_0^j\left(e^{-t_{\rm{rec}}/\tau_2}
                + B_2\right) + B_1^j e^{-t_{\rm{rec}}/\tau_1}
\label{e-Bfit}
\end{equation}
where $t_{\rm{rec}}$ is the time elapsed since the irradiation and
$j,\; (j=1,2,3)$ is the crystal index. This parametrisation yields,
for $t >> \tau_1$, a damage amplitudes ratio between crystals which is
independent from the elapsed time. The fits being almost
indistinguishable, the comparison between crystals remains independent
of the selected fit function.

A deeper insight into the damage amplitude ratio between positive
pions and protons was gained in \cite{r-PIONNIM} by looking at the
induced absorption profile along the depth of the crystals.  The
transverse induced absorption coefficients $\mu_{IND}^{TT}(420\;
\mathrm{nm})$ were obtained from measurements of the Transverse Light
Transmission (TT) performed sideways at various positions along the
crystals length, as indicated in figure \ref{f-SKETCH}. The profiles
for this quantity are plotted in figure \ref{f-TT}.

It was advocated in \cite{r-LTNIM} that damage could be proportional
to the density of stars, i.e. of inelastic hadronic interactions
caused by a projectile above a given threshold energy.  To verify this
hypothesis, the damage profiles from figure \ref{f-TT} have been
rescaled first of all to an identical fluence value, of
$10^{13}\mathrm{cm}^{-2}$, as it is justified due to the cumulative
nature of the damage.  The so-obtained profiles are shown in figure
\ref{f-NORMMU}.  In figure \ref{f-STARRHO}, the star densities from
FLUKA Monte Carlo simulations are shown as a function of depth in the
crystal, where arrows indicate the lengths covered by the crystals
during irradiation. One notices the fast drop, due to absorption, for
200 MeV $\pi^+$, while the density rises and stays high for 24 GeV/c
protons throughout the depth corresponding to $w1$ and $w3$.  The
comparison of the damage profile from protons in {\em w1} and {\em w3}
and pions in {\em w2} from figure \ref{f-NORMMU} with the simulated
star densities profiles in figure \ref{f-STARRHO} shows an agreement
that confirms in a striking way the hypothesised mechanism.  A direct
comparison is then performed in figure \ref{f-RATIOS}.  The ratios of
rescaled transverse absorption coefficients measured for positive
pions in {\em w2} and for protons in crystal {\em w1} are plotted as a
function of depth. The corresponding ratio of star densities obtained
from the FLUKA simulations described in \cite{r-LTNIM} and
\cite{r-HUH} is also plotted.  The measured ratios and the star
densities ratios are in agreement throughout, within the experimental
uncertainties.

This comparison establishes that the damage amplitudes caused by 24
GeV/c protons and 290 MeV/c positive pions in lead tungstate scale
according to the star densities caused by those particles in the
crystal as calculated by Monte Carlo simulations.
\section{Conclusions and outlook}
\label{s-CON}
Irradiation studies with 20 to 24 GeV/c protons and $^{60}$Co photons
in a fluence regime as expected at the LHC demonstrate that hadrons
cause a specific, cumulative damage in lead tungstate, which solely
affects the crystal light transmission while its scintillation
properties remain unaffected. Therefore, it is possible to monitor the
damage through light injection. All characteristics of the damage are
consistent with it being due to an intense local energy deposition
from heavy fragments. Since this damage mechanism should be absent in
crystals with elements below Z=71 \cite{r-FISS}, a hadron damage test
in such crystals should confirm the present understanding.  A positive
pion irradiation shows that the damage amplitudes caused by 24 GeV/c
protons and 290 MeV/c pions in lead tungstate scale like the star
densities those particles cause in the crystals according to
simulations. The results can thus be used to estimate the expected
damage for different experimental conditions.

\end{document}